\documentstyle[11pt,aaspp4]{article}
\begin{document}
\tighten
\def\secpoint{.\kern-.3em''\kern-.1em}
\def\uchii {UCH~{\sc ii}\ }
\def\hii{H~{\sc ii}\ }
\def\tablevspace#1{\noalign{\vskip#1}}
\def\tablebreak{\pt@line\pt@nlines\advance\pt@line by\m@ne\pt@nl}
\def\tabledoubleline{\tablevspace{8pt}\hline\tablevspace{3pt}\hline
        \tablevspace{8pt}}
\def\begintableline{\hline\tablevspace{8pt}}
\def\endtableline{\tablevspace{8pt}\hline}
\singlespace
%2345678901234567890123456789012345678901234567890123456789012345678901234567890
\title{Expansion of W~3(OH)}
\author{Jonathan H. Kawamura\altaffilmark{1} \& Colin R. Masson\altaffilmark{2}}
\affil{Center for Astrophysics, 60 Garden St., Cambridge MA 02138}
\altaffiltext{1}{present address: Division of Physics, Mathematics and 
Astronomy, California Institute of Technology, 320-47, Pasadena CA 91125;
jhk@caltech.edu}
\altaffiltext{2}{present address: Renaissance Technologies Corp., 600 Route
25A, East Setauket, NY 11733; colin@rentec.com}
\begin{abstract}

A direct measurement of the expansion of W~3(OH) is made by comparing
Very Large Array images taken $\sim 10\,\rm yr$ apart. The expansion is
anisotropic with a typical speed of $\rm 3\,\,to\,\, 5\, km\,s^{-1}$,
indicating a dynamical age of only $2300\,\rm yr$. These observations
are inconsistent with either the freely expanding shell model
or a simple bow shock model. The most favored model is a slowly 
expanding shell-like \hii region, with either a fast rarefied flow or
another less massive diffuse ionized region moving towards the
observer. There is also a rapidly evolving source near the projected
center of emission, perhaps related to the central star.

\end{abstract}


\section{Introduction}

Ultra-compact \hii (UCH\ {\sc ii}) regions are small but dense envelopes 
of ionized gas that form out of the natal material around early-type 
massive stars that have just formed. While the general scenario of 
how massive stars form and how they might appear were theoretically
described more than several decades ago (Davidson \& Harwit 1967, 
Larson 1969a, b, 1972, Yorke \& Kr\"ugel, 1977), the lack of observational 
probes have frustrated efforts to understand these regions in greater detail.
One of the pressing questions raised recently regarding high-mass 
star formation is exactly how \uchii regions evolve with time. 
Specifically, how long \uchii regions persist is currently a puzzle.

\uchii regions are typically under 0.1 pc in size, 
$\sim 10^{17}\,\rm cm$, but extremely dense, $n_e>10^5\,\rm cm^{-3}$. 
The high density and temperatures of $\sim 10^4\,\rm K$, typical of 
\hii regions, together suggest that there should be strong pressure 
imbalance between the ionized region and its surrounding medium. 
Massive molecular outflows are commonly observed near \uchii regions, 
and highly turbulent velocities are also present in spectral 
line profiles. \uchii regions are also closely 
associated with sites of very energetic $\rm H_2O$ and OH masers, 
indicating that there is a strong interaction between the central stars,
which supply the energy for the masers, and the molecular material outside 
the ionized region. Massive stars are also known to possess
strong stellar winds that greatly affect their environment.
This combination of observational clues suggests that the ionized
regions should be rapidly both evolving and expanding, provided
there is no constraint to the expansion. For an unconstrained
expansion the estimated age of \uchii regions is approximately
the size of the region divided by the sound crossing speed, and for 
most \uchii regions the inferred age is $\sim 10^3$ to $10^4\,\rm yr$. 
Presumably, after this phase, most of the material that shrouds the 
star is blown away, revealing optically the star with its \hii region.

However, several studies of the statistics of \uchii regions
suggest that the lifetime of the ultra-compact phase is of order
$10^5\,\rm yr$, considerably longer than the dynamical age. 
Nearly two decades ago, Habing~\& Israel~(1979) recognized that
the time scales of several concurrent phenomena were inconsistent with
the dynamical age of \uchii regions. The time scale for natal 
material to coalesce to form stars, and the time over which $\rm H_2O$ 
masers are thought to exist are both about $10^5\,\rm yr$. Since 
both processes are seen in vicinity of almost all \uchii 
regions, it is unlikely that the \uchii phase could be significantly 
shorter in duration than the two processes. Therefore, \uchii regions
are perhaps longer lived than suggested by the dynamical age.

More recently the analysis of Infrared Astronomy Satellite (IRAS) sources by 
Wood \& Churchwell~(1989b) showed that between perhaps 10 and 20\% of all 
O stars are still embedded in molecular clouds. This implies that 
the \uchii phase lasts for up to 20\% of the lifetime of an O star or
$\sim 10^5\,\rm yr$, in good agreement with the age predicted by
Habing \& Israel~(1979). Probably the most convincing argument for 
\uchii regions having lifetimes longer than their dynamical ages is the
follow-up Very-Large-Array~(VLA) survey of \uchii regions by 
Wood \& Churchwell~(1989a). They detected a large excess
in the number of \uchii regions over that expected if the number of
\uchii regions were simply given by the total galactic O star population
multiplied by the ratio of the dynamic age to the lifetime of an O star.

Thus, either \uchii regions are presently too numerous in the 
Galaxy or their lifetimes are longer than expected by a factor 
close to $10^2$. Discounting the former argument, Wood \& 
Churchwell~(1989a) concluded that the \uchii phase must persist 
longer than the dynamical age. 
This conclusion has stirred much theoretical interest and spurred 
many observational projects (e.g. Van Buren et al. 1990, 
Mac Low et al. 1990, Hollenbach et al. 1994, Garc\'\i a \& Franco 1996,
Akeson \& Carlstrom 1996).
It is thought that either \uchii regions are effectively and 
stably confined to prevent their rapid expansion, or that the observed 
size of the emission structure does not change because the material is 
constantly being replenished. The first reason may apply for some
\uchii objects with shell-like morphology, while the second may
apply for others with core-halo morphology, and a combination has
been suggested in a model for cometary objects as bow shocks around
stars moving through molecular clouds (Van Buren et al 1990).

Clearly, one of the obstacles to understanding \uchii regions is 
the lack of good observational probes. First, they are totally obscured
at optical wavelengths. The central star exciting the \hii region
W~3(OH) is obscured by dust with an V-magnitude extinction of
$\sim 50$ (Wynn-Williams, Becklin \& Neugebauer 1972). Presently, \uchii 
regions can be best studied at centimeter 
wavelengths using synthesis imaging. Observations in the far infrared 
can measure the flux from the dusty cocoon that surrounds the \hii region 
revealing the bolometric luminosity, which indicates the star-type. 
Millimeter-wave interferometers do not as yet resolve the regions 
adequately, but should be an excellent probe of the molecular material
near the \hii region if slightly higher resolutions are attained.

The compact \hii region W~3(OH) is a limb-brightened shell of dense ionized
gas around an O7 star that has recently formed (Dreher \& Welch~1981).
Regarded as the prototypical \uchii region, W~3(OH) is associated
with a prominent cluster of OH masers (Norris \& Booth 1981). The central star
is totally optically obscured by an extended dusty cocoon that
envelops the \hii region (Wynn-Williams et al. 1972).
The \hii region is ensconced in a massive molecular cloud,
$M\sim 2000\,M_\odot$, which extends about 1~pc across, and exists
in the vicinity of two core components (Wilson, Johnston, \&
Mauersberger 1991). The more massive component, with mass $M\sim 60\,M_\odot$,
is associated with the source identified by Turner \& Welch (1984),
which is associated with the cluster of $\rm H_2O$ masers about
$6''$ east of the main \hii region. W~3(OH) is associated with
the less massive component, with mass $M\sim 10\,M_\odot$.
Presumably, a significant fraction of the mass in the
protostellar core condensed to form the $\sim 30\,M_\odot$ star
that excites the \hii region. The ionized region itself is
a bright source at centimeter wavelengths, with a total flux of 
$2.2\,\rm Jy$ at 15~GHz, which is approximately the turnover frequency. 
The shell measures approximately $1''$ across, and for a distance of 
2.2~kpc (Humphreys 1978), the \hii region is only 
$3\times 10^{16}\,\rm cm$ in diameter. With this flux
density and dimensions for the nebula, the emission measure is in
excess of $10^9\,\rm pc\,cm^{-6}$. The dynamical age of the \hii 
region, given by its diameter divided by the plasma sound speed 
($c\approx 10\,\rm km\,s^{-1}$), is only $\tau_{dyn}=10^3\,\rm yr$. 

This paper describes direct observations of the changes in the W~3(OH)
region by comparing centimeter synthesis images made at several time epochs. 
Specifically, we exploit the proven technique of difference mapping using 
VLA images taken at epochs separated by several years to observe this
well-studied bright \uchii region. This robust method 
has been used to measure accurate distances to ionization-bounded 
planetary nebulae (Masson 1986). In this application the distance is 
measured by combining the angular expansion with spectroscopically 
determined velocities. In the present paper, the goal is not to measure 
the distance to the \hii region, but rather to use the difference maps to 
understand its evolution.

\section{Observations} 

All observations were made using the NRAO
\footnote{The National Radio Astronomy Observatory is a facility of the
National Science Foundation operated under cooperative agreement by
Associated Universities, Inc.}
VLA synthesis telescope in the A-configuration, observing in the 
continuum mode at 2~cm. There are three sets of observations for 
W~3(OH): 6 Mar 1986, 9 Mar 1990 and 25 Jun 1995. 
The observations are long tracks to ensure nearly complete, uniform spatial 
coverage of the source. Thus, the spatial coverage is similar in 
all epochs. Additionally, a small amount of time in the 
1986 run was devoted to making 1.3~cm observations to aid in the 
modeling. The amplitude was calibrated with 3C~286 in the 
1986 and 1990 observations, and with 3C~48 in the 1995 
observations. The flux densities were assumed to be 3.49~Jy for 3C~286 
and 1.81~Jy for 3C~48. For all epochs the phase calibrator was 0224+671.
The 2~cm flux density for the phase calibrator 0224+671 was 0.89~Jy, 
1.70~Jy, and 2.73~Jy in 1986, 1990 and 1995, respectively. 
To improve the sensitivity the intermediate frequency bands were 
averaged when appropriate. Each data set was calibrated separately 
in order to make independent measurements of the flux density. 
We estimate that the uncertainty of the amplitude of the data is 
5\% at 2~cm, and 10\% at 1.3~cm. The flux densities and a summary of the 
observations are listed in Table \ref{hii_observations}. The
final 1.3~cm map was made with several iterations of self-calibration
and CLEAN, convolved with a $0\secpoint 1$ circular Gaussian beam.
A contour map of the region is shown in Figure~1. Contour maps of the 
main \hii region at 15~GHz and 22~GHz are shown in Figure~2.

\begin{table}[t]
\begin{center}
\begin{tabular}[t]{clccc}
\begintableline
\ &     & No. of        & Total         & RMS\\
$\lambda$ & Date        &  Visibilities & Flux Density  & Noise \\
(cm)& & ($\times 10^3$) & (Jy) & ($\mu\rm Jy/beam$) \\
\tabledoubleline
2
& 1986 Mar 6
& 434
& $2.23\pm 0.11$
& 65 \\
\
& 1990 Mar 9
& 302
& $2.05\pm 0.10$
& 45 \\
\
& 1995 Jun 25
& 881
& $2.35\pm 0.12$
& 40 \\
\\
1.3
& 1986 Mar 6
& 184
& $2.8\pm 0.2$
& 490 \\
\endtableline
\end{tabular}
\end{center}
\caption{Observation Parameters}
\label{hii_observations}
\end{table}

\begin{figure}[tf]
\vspace{3.5in}
\centerline{
\includegraphics{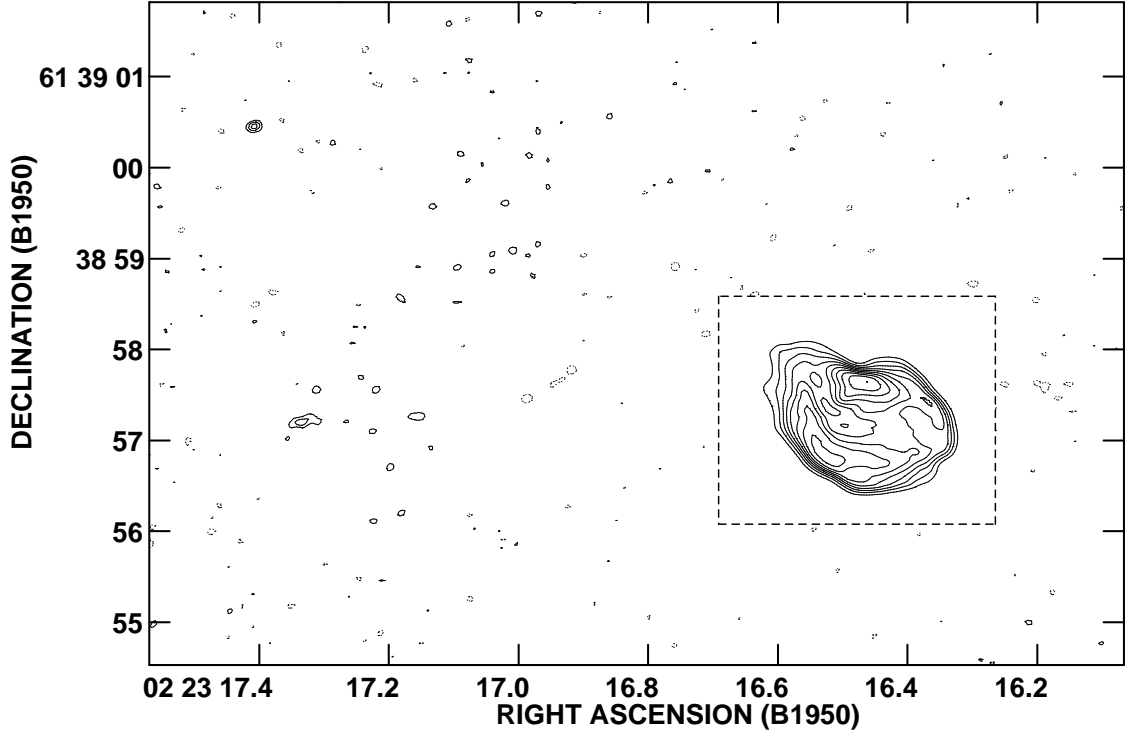}
}
\vspace{0.5in}
\caption{ A contour map of the W 3(OH) region observed at 15~GHz in 1986.
The main source is enclosed by the dashed box, in which the contour
levels are 10\% of the peak brightness, or $3.2\,\rm mJy\,beam^{-1}$.
Outside the dashed box the contour levels are integer multiples
of $200\,\mu\rm Jy\,beam^{-1}$. The source associated with water
masers is $6''$ directly E of the main source; there is another
source $8''$ to the NE.}
\end{figure}

\begin{figure}[t]
\vspace{3.5in}
\centerline{
\includegraphics{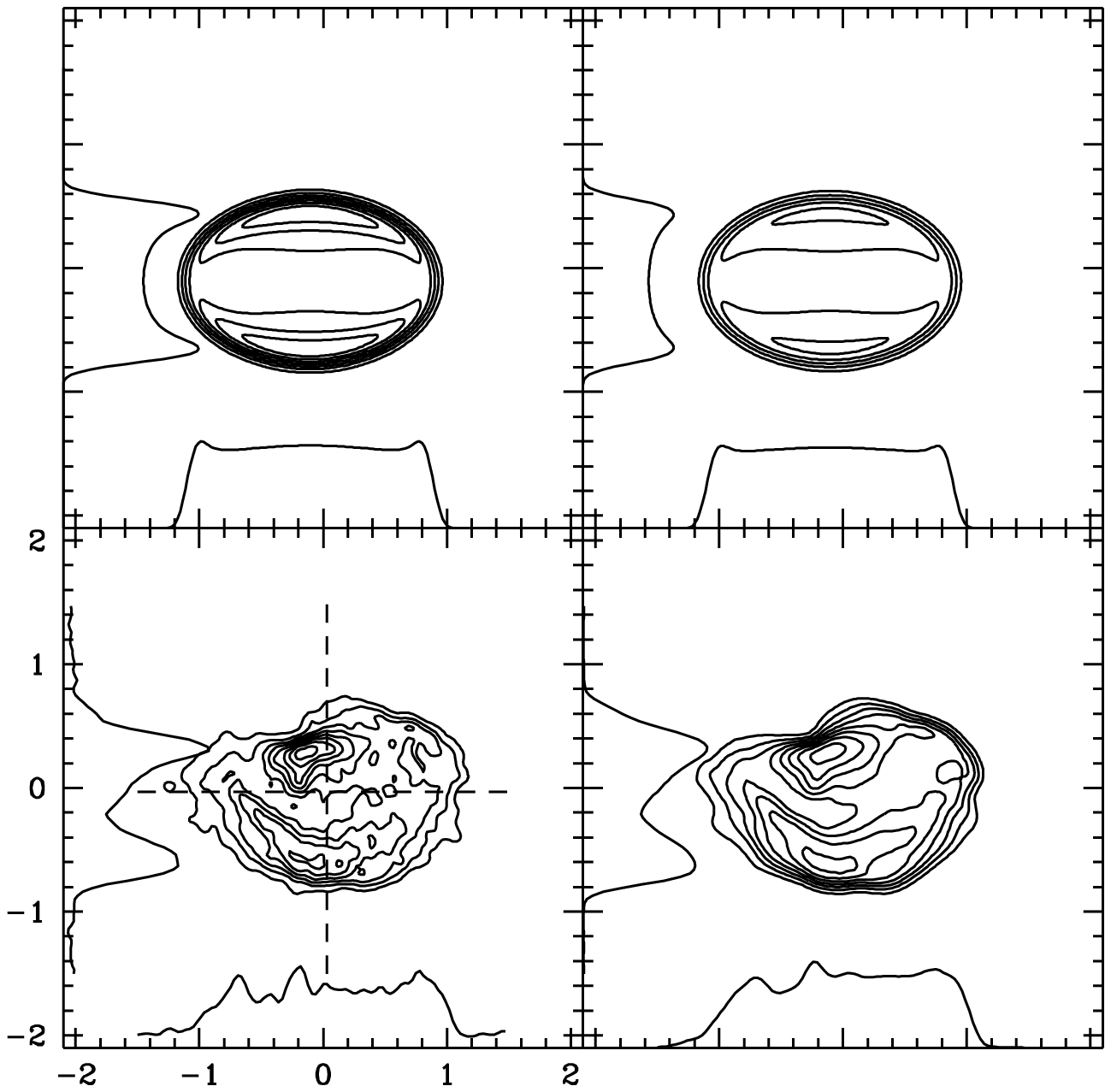}
}
\vspace{0.5in}
\caption{Comparison of model calculations and actual maps.
The bottom panels show the contour maps of W~3(OH), which, for
the purposes of presentation, have been rotated by $30^\circ$ in
position angle. The figure on the lower left is a 22 GHz map, and lower
left a 15 GHz map. The top panels are contour maps of the
model. Profiles of the brightness along the major and minor
axes of the \hii region are also plotted in the figure.
The scale is in asec.}
\end{figure}

The cross-calibrated difference mapping technique described by
Masson~(1986) is used to generate difference maps between epochs.
In this method the data from each epoch is calibrated and imaged
in the standard manner to select the epoch with the best data.
The image from the best epoch is then used as the model
in the cross-calibration of the data from the other epochs.
This ensures that all epochs have common phase errors across
the visibility plane, which would otherwise be manifest as systematic
features in difference maps. This step also conveniently aligns
the position of the image of subsequent subtraction. The cross-calibrated 
data is imaged, and its CLEAN components are subtracted from the best data 
set in the visibility domain. The resultant data is imaged and CLEANed.
All images are then restored with the same $0\secpoint 128$ circular
Gaussian beam. 

The difference mapping technique is particularly suited to quantify
the angular expansion of objects with sharp boundaries, such as 
ionization-bounded nebulae like young planetary nebulae and W~3(OH). 
The angular movement is calculated by either directly computing
the ratio of the difference signal and the gradient of the emission,
or by suitably modeling the movement. Obviously, both methods should 
lead to results which concur. In the case of a reasonably uniform
expansion, there is little ambiguity about the meaning of the
difference map measurements. The technique can also be applied to
measure proper motion if there is a fiducial point source in the 
field. 

The main difficulty in the subtraction step is the uncertainty in the
source flux density. The difference maps may contain signals
only a few times the map noise. For W~3(OH) this means that in order
to detect a signal in the difference map, the flux must be
known to about 1\%. This level of accuracy in the amplitude 
is not readily possible with the VLA, and at 15~GHz, an uncertainty
of $\sim5\%$ is typical. A possible way to sidestep the
measurement uncertainty is to assume that the flux density remains
constant. This assumption, for example, is reasonable in the case
of optically thin emission from a planetary nebula. In the 
subtraction step for W~3(OH) the total fluxes were assumed 
to be same for all epochs. 

\section{Results}

Because there are three sets of observations, three difference maps 
were also generated. The difference maps are shown in the upper
panels of Figures 3, 4 and 5, in which the difference maps are
shown in contour superimposed on a grey-scale image of the region
itself. Figure~3 is a difference map made from observations
separated by 9.3~yr, Figure 4 by 5.3~yr, and Figure 5, by 4.0~yr.
Since the difference maps were constructed by subtracting
earlier from later data, the expansion is seen as an incomplete
positive ring of flux surrounding a negative region. The assumption
that the total flux remains constant forces the difference maps
to have zero integrated flux and as a result the region inside
the ring has an average negative value. Having a number of 
difference maps is important and convenient because it allows us to 
easily identify spurious signals. It is also useful to compare maps 
generated from observations separated by increasing time baselines to 
recognize and identify trends. If there are complicated difference signals, 
then a number of different difference maps obviously makes it easier 
to interpret their meaning. Most notably, in this sequence of difference 
maps the brightness of the ring is clearly related to the time interval 
between observations.

\begin{figure}[p]
\vspace{6in}
\centerline{
\includegraphics{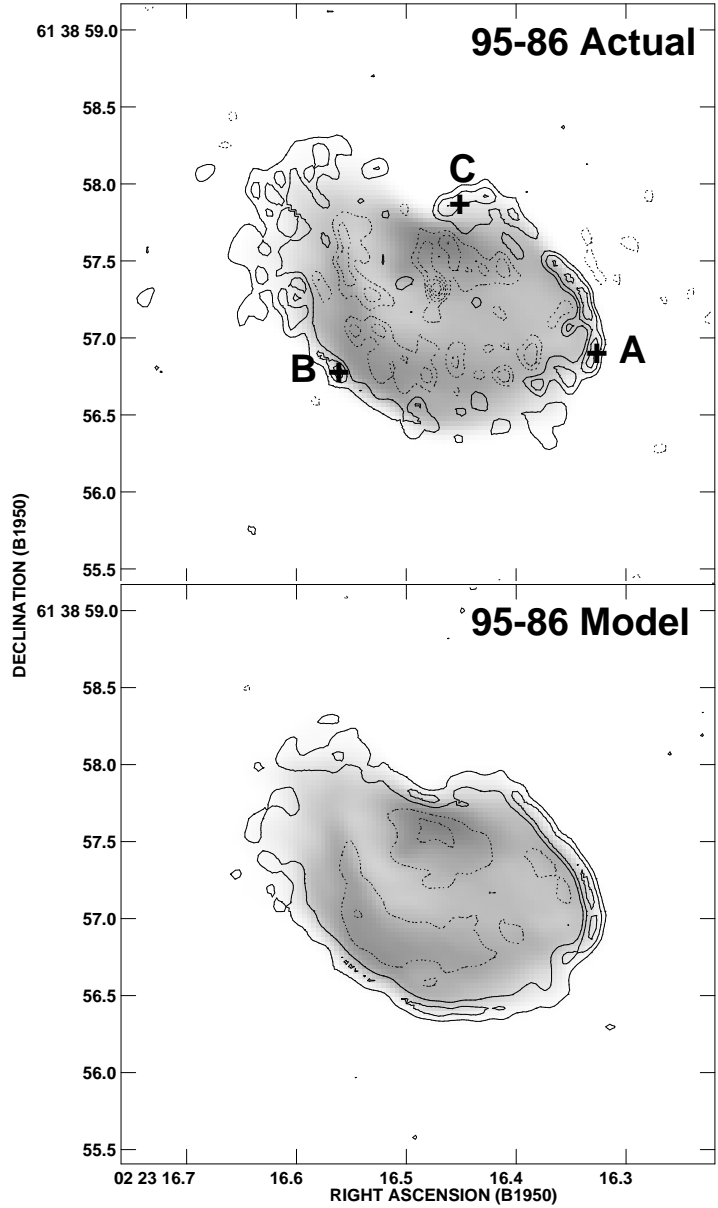}
}
\vspace{.5in}
\caption{Difference map of W~3(OH) with a time baseline of 9.3~yr.
The 15~GHz image is shown in grey scale, and the difference map is shown
in contour. In the top panel the actual $95-86$ difference map
is shown. A difference map generated from subtracting the 1986
image from a simulated self-similar expansion of the 1986 data
is shown in the bottom panel. The frequency of the visibility
was divided by a factor of 1.004, which corresponds to an expansion
in the image plane. The contour levels are $200\,\mu\rm Jy\,beam^{-1}$
in both panels. Note the presense of a $-1\,\rm mJy$ signal near
the center of the nebula in the upper panel. 
}
\newpage
\end{figure}

\begin{figure}[p]
\vspace{6in}
\centerline{
\includegraphics{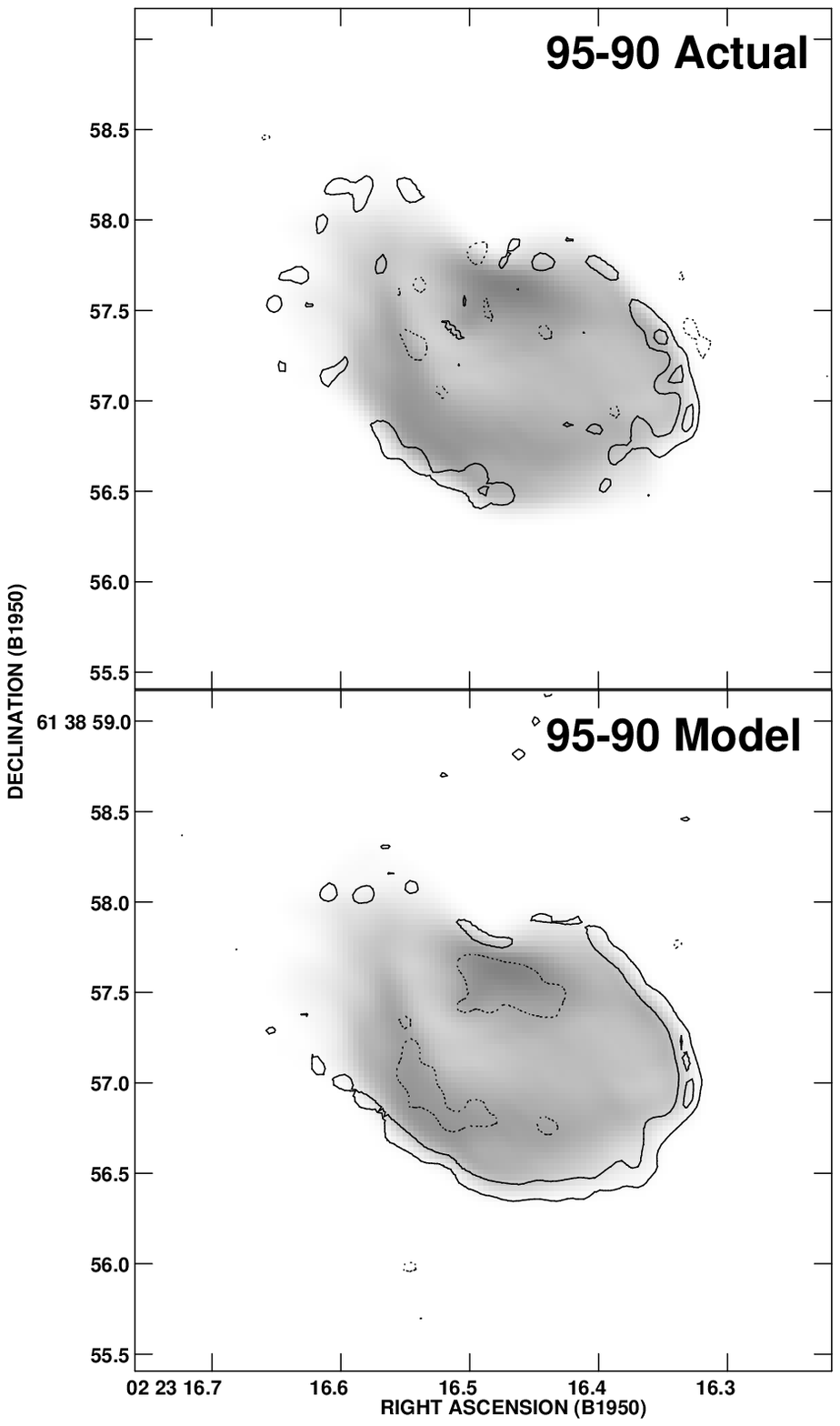}
}
\vspace{0.5in}
\caption{
Difference maps of W~3(OH) with time baseline of 5.3~yr.
The 15~GHz image is shown in grey scale, and the difference map is shown
in contour. In the top panel, the actual $95-90$ difference map
is shown. The simulated difference maps is shown in the lower
panel, in which the frequency of the visibility data was
divided by a factor of 1.0023. The contour levels are
$200\,\mu\rm Jy\,beam^{-1}$ in both panels. 
}
\end{figure}

\begin{figure}[p]
\vspace{6in}
\centerline{
\includegraphics{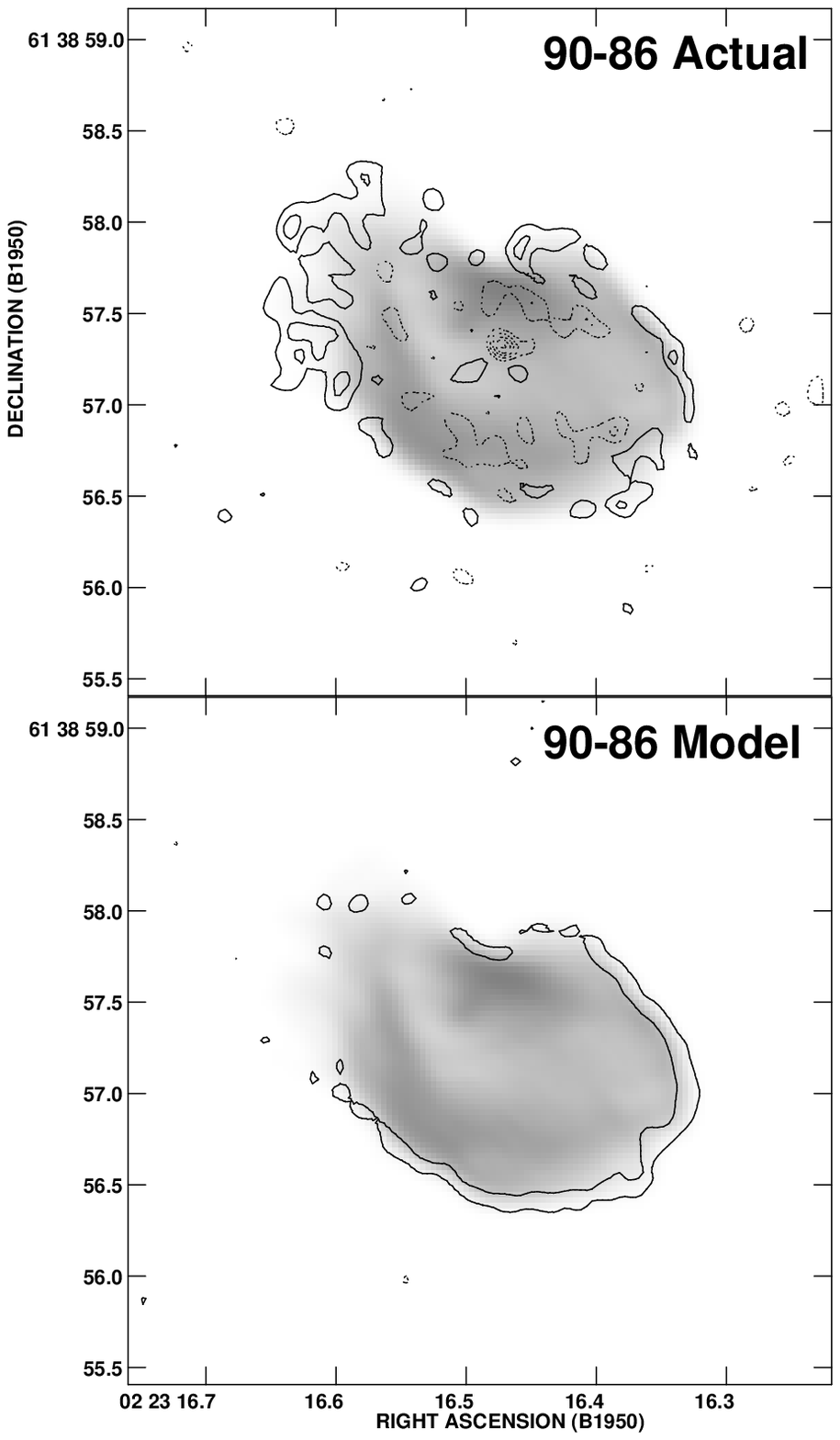}
}
\vspace{0.5in}
\caption{Difference maps of W~3(OH) with time baseline of 4.0~yr.
The 15~GHz image is shown in grey scale, and the difference map is shown
in contour. In the top panel, the actual $90-86$ difference map
is shown. The simulated difference maps is shown in the lower
panel, in which the frequency of the visibility data was
divided by a factor of 1.0017. The contour levels are
$200\,\mu\rm Jy\,beam^{-1}$ in both panels. Note again the
presence of the negative feature in the upper panel. 
}
\end{figure}

The difference maps reveal not only the expansion of W~3(OH), but
also evidence of variability among other sources in the region.
The first of these is located at the projected center of the 
main component, indicated in Figure 3. In fact, the feature is 
located about $0.05''$ S of the centroid of the emission, and could 
be directly related to the exciting star. The second source is the 
object located about $7''$ NE of the main source. The third source is 
the object associated with the $\rm H_2O$ maser complex, thought to be 
a source of synchrotron emission arising from protostellar 
jet (Reid et al 1995). 

\subsection{Measuring the expansion}

Because the expansion between any two epochs is small compared
with the beam size of our observations, the angular 
expansion rate may be computed from the difference signal
by $\dot\theta=\Delta f(r)/f'(r)t,$
where $t$ is the time interval, $\Delta f(r)$ is the signal in the
difference map, and $f'(r)$ is the observed gradient in the emission 
along the line of movement. The angular proper motions were 
calculated at three positions
in the difference maps, marked as `A',`B', and `C' in 
Figure~3, and are listed in Table \ref{hii_proper_motion}.  
The profiles of the source and difference map intensities along
radial cuts centered near these positions are shown in Figure~6. 
Our experiments with the data show that the signals in the difference maps
are repeatable to about $\sim 0.2\,\rm mJy\,beam^{-1}$ among
different trial reductions. This level of uncertainty is, furthermore,
equivalent to a flux uncertainty of about 1\%. We therefore
adopt this value as the uncertainty of the difference signal, from
which the errors in Table \ref{hii_proper_motion} were calculated.
This uncertainty dominates all other sources of errors, including 
map noise.

\begin{figure}[tf]
\vspace{3.9in}
\centerline{
\includegraphics{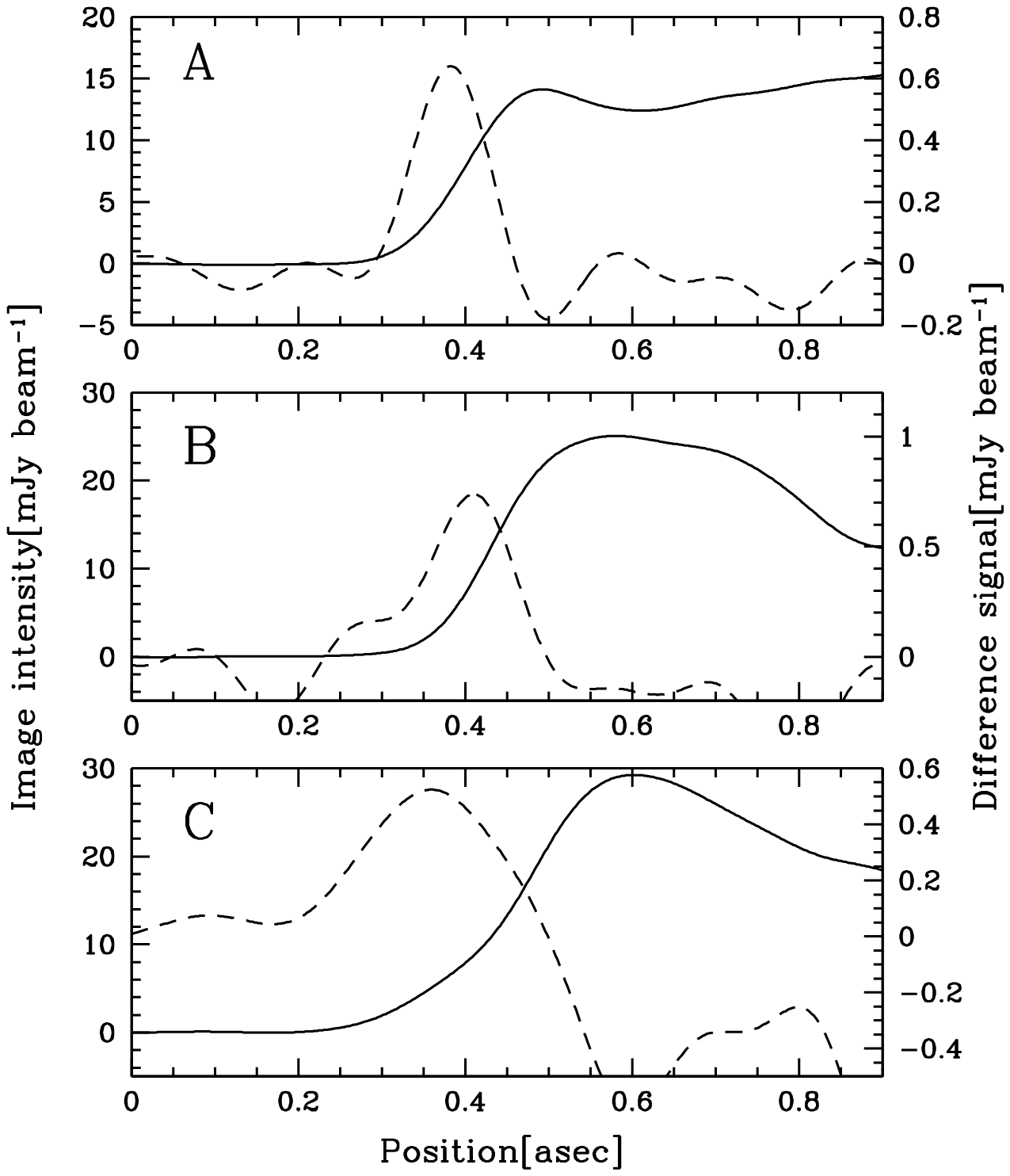}
}
\vspace{0.5in}
\caption{ Profiles of the source and difference map intensities
along radial cuts centered at positions A, B and C. The solid
lines represent the source intensity, and the dashed line
the difference map intensity.
}
\end{figure}

\begin{table}[t]
\begin{center}
\begin{tabular}[t]{cccc}
\begintableline
Difference
& &
Position
\\
Map
& A
& B
& C
\\
\tabledoubleline
$95-86$
& $0.7\pm 0.2$
& $0.8\pm 0.4$
& $0.3\pm 0.5$
\\
$95-90$
& $0.4\pm 0.1$
& $0.4\pm 0.2$
& $0.2\pm 0.4$
\\
$90-86$
& $0.8\pm 0.3$
& $0.6\pm 0.6$
& $0.7\pm 0.7$
\\
model
& $0.5\pm 0.1$
& $0.4\pm 0.1$
& $0.2\pm 0.1$
\\
\endtableline
\end{tabular}
\end{center}
\caption{Proper motion measurements of W~3(OH). ($\rm m.a.s.\,yr^{-1}$)}
\label{hii_proper_motion}
\end{table}

When the angular proper motion is combined with the distance to
the region, an expansion speed may be calculated. The proper motions are 
calculated by assuming a distance of 2.2~kpc. If we average the
angular proper motions calculated at each position, then the
expansion speed at position A is $v\approx 5\rm\,km\,s^{-1}$, 
at position B is $v\approx 5\rm\,km\,s^{-1}$, and at position C is
$v\approx 3\,\rm km\,s^{-1}$. We note that the 
typical expansion speed is strikingly similar to the non-parametric 
estimate of the OH maser expansion rate (Bloemhof, Reid \& Moran 1992).

\begin{figure}
\label{hii_slices}
\end{figure}

To test our analysis and to calibrate the expansion measurement,
we made an artificial dataset corresponding to a self-similar
expansion of the nebula. The difference maps resulting from this
simulated expansion of the nebula are shown in the lower panels of
Figures 3, 4 and 5.  Here, the frequency parameter in the 1986 
visibility data was divided by a factor $1+\epsilon$, 
where $\epsilon=(4.3\pm 1.0)\times 10^{-4}\,\rm yr^{-1}$.
The uncertainty given here corresponds to the range by which
$\epsilon$ can vary so that the resulting expansion signal 
changes by less than $0.2\,\rm mJy\,beam^{-1}$, which was adopted 
as the uncertainty above. Changing the frequency in this manner 
artificially contracts the visibility plane, which corresponds 
to an expansion in the image domain. Difference maps are 
generated using the same procedure as for making normal 
difference maps: the CLEAN components from the unaltered 
data are subtracted from the altered data set 
in the visibility domain, and the difference data
are imaged and CLEANed. The difference images made in this manner
show fairly good resemblance to the panels immediately above them.
This similarity strongly supports the idea that the expansion 
signatures in the real difference maps are indeed from 
actual expansion of the ionized region. The angular expansion
rates calculated from using $\epsilon$ are also listed in 
Table~\ref{hii_proper_motion}.

One salient feature in the simulated difference maps is that the positive
expansion signature forms a half circle around the source as in the lower
panel of Figure~3.
This is caused by the simple fact that difference maps are most sensitive
to movements that occur along a steep gradient in the image. It is 
important to realize that it is not caused by some anisotropy in the 
expansion. Clearly, difference maps are not sensitive to a motion 
that occurs transverse to a gradient. Towards the western limb
of W~3(OH) the ionization front forms a very sharp cut-off in the
emission, and thus there is a positive feature around the limb.
Towards the east, however, there is some extended emission
and there is less signal in the simulated difference maps.
The presence of some significant positive features 
towards the eastern edge of the source in the actual difference 
maps suggest that the material there is moving very rapidly.
This is a tantalizing possibility because such a rapid 
flow might explain the tail that extends to the NE
and is very prominent in maps made at longer wavelengths.
Also, it may explain the velocity gradient of the recombination
line seen across the \hii region. It is also quite 
possible that these changes simply reflect changing physical 
conditions in the region that occur without any motion of 
the material. However, since positive features appear in those 
difference maps that involve the 1986 epoch and not in 
the 1995--1990 map, it is most likely that these are merely 
residual calibration errors in the 1986 map. 

A major strength of the difference mapping technique applied to
objects undergoing expansion is that it is possible to assign a dynamical age 
without making any assumptions about the physical model. The explicit
method is to divide the angular size by the angular proper motion.
An equivalent and more convenient method is to calculate the
age using the factor used in the simulated expansion, 
${\rm age} \approx 1/\epsilon=(2300\pm 600)\,\rm yr$.

\subsection{Variability in the central region}

At the projected center of the \hii region in the difference maps,
there is a negative feature in the 1995--1986 (Figure 3) and 1990--1986
(Figure 5) maps. This feature is not readily visible in the images of 
the \hii region and only apparent in the difference maps.
Its position coincides with the centroid of the emission of the 
entire \hii region to within $\sim 0\secpoint 05$, and may therefore
be related to the central star. This feature is unresolved, 
and its strength is $\Delta f_{\nu} \sim -1\,\rm mJy$, with a 
peak brightness temperature of $\Delta T\sim -1000\,\rm K$. 
On either side of the feature, along the E-W direction, there are
positive features, which may indicate a mass flow. Because the
source is unresolved, an upper limit to the physical size is 
$<300\,\rm AU$. 

Incidentally, Dreher \& Welch~(1981) described a feature in the center 
of their 1.3~cm map, which they tentatively ascribed to emission from 
residual material in free-fall in the cavity. Under such a configuration,
the density varies as $r^{-3/2}$, and the material becomes detectable 
only very near the star. In our new maps such a structure is not seen. 
While it is possible that the feature became undetectable when our 
observations were made, the maps made by Dreher \& Welch seem to 
be poorly calibrated.

\subsection{Variability of other sources in the region}

A source $6''$ to the east of W~3(OH) has recently come under
great scrutiny (Turner \& Welch, 1984). While at centimeter-wavelengths 
the source is rather weak, its spectrum was discovered to be consistent with
non-thermal synchrotron emission (Reid et al. 1995). Further observations
with millimeter-wave interferometers have shown that its 
thermal continuum emission greatly dwarfs that of the main source
(Wink et al. 1994; Wilner, Plambeck \& Welch 1996). The object
has been proposed to harbor a massive star in its earliest stage
of formation. 

The difference maps of the region are shown in Figure~7.
Reid et al. (1995) used the flux values averaged from 1986 and 1990
data, but the variability of the flux is about $+30\,\mu\rm Jy\,yr^{-1}$. 
The object is barely resolved and is elongated in the E-W direction, as 
noted in Reid et al. (1995). Furthermore, since the signal in the
difference map is slightly to the E of the centroid of the emission, 
the source is probably getting further elongated. It is tempting
to convert this difference signal to a velocity of a `flow,' but
there is very little to support that it represents any movement
of material. The time variability of this object suggests that the 
spectrum, which was taken during observations separated by up to 5~years, 
might be incorrect. A simultaneous measurement of the flux at each frequency
is necessary to determine unambiguously the spectrum of the object. 

\begin{figure}[tf]
\vspace{4.1in}
\centerline{
\includegraphics{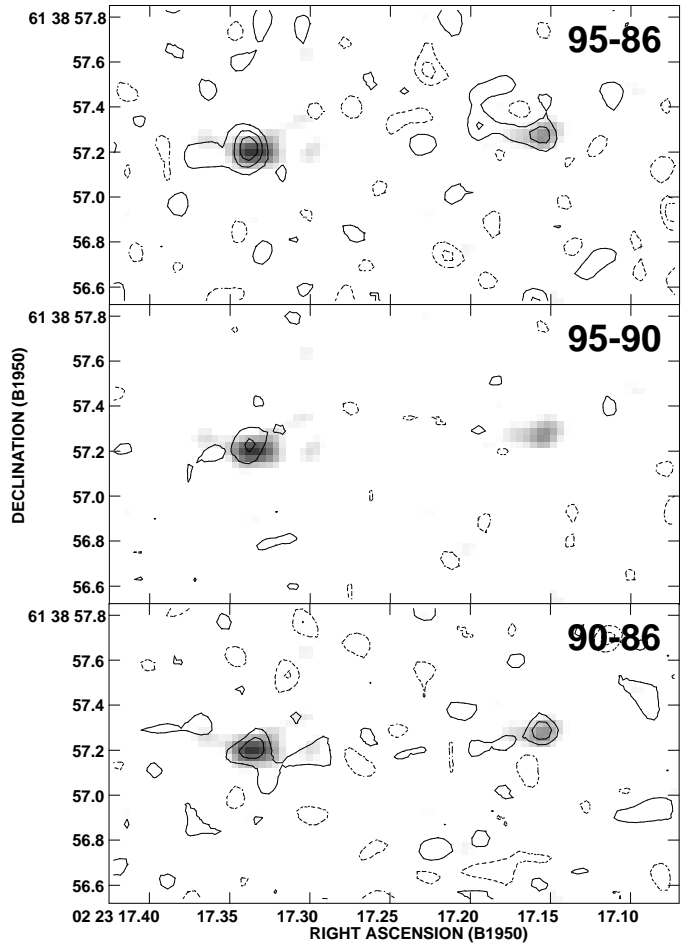}
}
\vspace{0.5in}
\caption{  Difference map of the region around the synchrotron
source, about $6''$ E of W~3(OH). The difference maps are
shown in contour, and the image of the region is shown in gray-scale.
The object to the left is the synchroton source, and the
source to the right is believed to be its associated
jet. The contour levels are $100\,\mu\rm Jy\,beam^{-1}$.
}
\end{figure}

There is another feature which was present in the first
epoch map but disappeared in those from later observations. The 
source, shown in Figure~8, is located about $7''$ NE of 
the main source, and is nearly unresolved. We identify this source with the 
`new source' described in Baudry et al. (1993), who measured a 
flux density of 0.8~mJy and a peak brightness temperature of 750~K at 
8.1~GHz. There is a nearby weak 1.6~GHz OH maser, which is probably
excited by the source. In 1986 the source had a 15~GHz flux density of 
$1.0\,\rm mJy$, and then the source went below the detection threshold
in 1990 and 1995. At the later epochs, we can place an upper limit
of $0.1\,\rm mJy$ to the flux density. 

\begin{figure}[p]
\vspace{4.5in}
\centerline{
\includegraphics{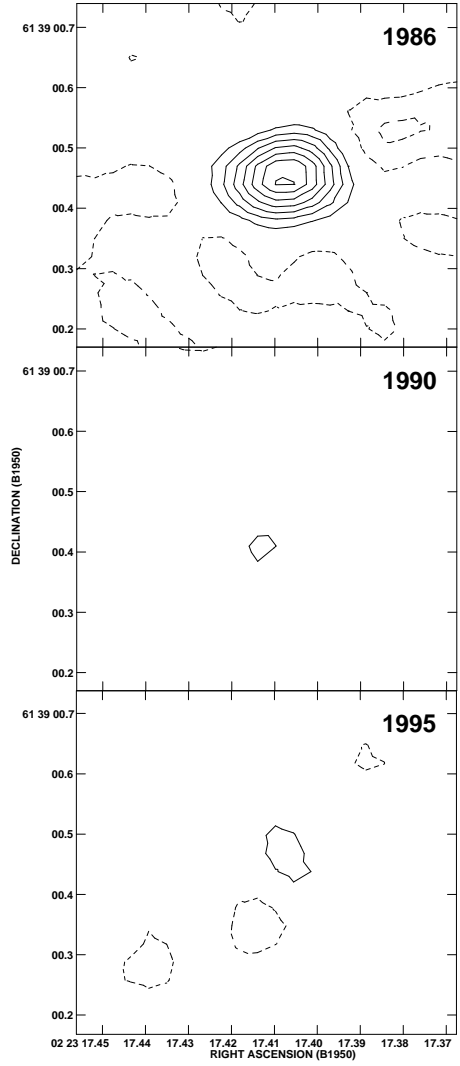}
}
\vspace{0.5in}
\caption{   A source $7''$ NE of the main source first
described by Baudry et al (1993). The top panel is a map
from the 1986 observations; middle, 1990; and bottom, 1995.
The contour levels are $100\,\mu\rm Jy\,beam^{-1}$.
}
\end{figure}

Baudry et al. (1993) observed this source in March, 1990, 
nearly at the same time as the 1990 epoch observations.
If we assume that we have properly identified the source,
there are three possibilities of what happened. 
The first one, which we feel is most likely,
is that the source had already ``turned off'' before
March 1990, and has a negative spectral index, making this object 
perhaps similar to the $\rm H_2O$ maser source immediately
to its south. The other possibility is that the source ``turned off'' 
between Baudry's March 1990 and our March 1990 observations, which would 
mean that we cannot infer any information about its spectral index. 
This possibility is not unphysical, since there is
sufficient time between the observations for such a change to 
occur. Finally, it is also possible that this source is cosmological, 
and completely unrelated to the molecular cloud complex. However, 
Baudry et al. (1993) calculated that the
the number of detectable extragalactic sources of greater or same
brightness as the source, and concluded that such sources
would be exceedingly rare. Also, the vicinity of the OH maser
and the source suggests there is a relation between the two. 

\section{Discussion}
\subsection{A static model for W~3(OH)}
\label{modelsection}

In order to understand better the physical conditions in and around 
the \hii region, we have constructed a static model for the \hii region.
Dreher \& Welch (1981) attempted to model the 
\hii region as a spherical shell of ionized gas, but realized 
that this model could not reproduce the high
degree of limb brightening that is actually observed. Moreover,
rather than having a single model to describe the \hii region, 
the authors used different parameters to fit the brightness profiles 
along cuts made along the major and minor axes. The model 
is a spherical shell with inner and outer radii of 1300 AU and 1900 AU, 
respectively, with an electron density of $\sim 2\times 10^5\,\rm cm^{-3}$. 
They inferred an electron temperature of $9400\pm 1500\,\rm K$ by comparing 
the peak brightness in their 2~cm and 1.3~cm maps. Although the spherical 
models give a reasonable representation of the nebulae, the prolate 
ellipsoidal model, combined with the new observations, gives a better 
accounting for its structure and a better estimate of the density of the 
material in the shell.

From the present 2~cm and 1.3~cm maps, we can similarly measure the
electron temperature. The peak brightness temperature occurs near 
the northern limb and is $10700\pm 500\,\rm K$ at 15~GHz and 
$8900\pm 500\,\rm K$ at 22~GHz. With these numbers we infer an electron 
temperature of $11000\pm 1600\,\rm K$, which is somewhat higher than but 
within error of that measured by Dreher \& Welch (1981). However, there 
appears to be some non-uniformity of the temperature within the nebula, 
ranging from $\sim 9300\,\rm K$ near the western limb and $\sim 10^4\,\rm K$
near the southern limb. For the modeling described below, we
adopt an average electron temperature of $10^4\pm 1000\,\rm K$.

The prolate ellipsoidal model has been successful in accounting 
the appearance of a number of planetary nebulae with widely varying 
morphology, despite its simplicity (Masson~1986). 
The model assumes a shell of uniform density, 
whose thickness varies as the inverse square of the distance from 
the central star. This property approximately accounts for the 
inverse square dilution of the radiation field. In essence, it 
ensures that the number of ionized gas particles in any solid angle 
viewed from the star is the same regardless of the direction. 
In a spherical model high limb-brightening can only be obtained
by invoking a very thin shell. In our prolate model with
ionization constraints, the ends of the shells are
necessarily fainter than the equator, so lines of
sight which pass closer to the major axis show greater
degrees of limb brightening. The model has 6 parameters: 
opacity, minor and major axes of the ellipsoid, thickness of 
the shell at the minor (or major) axis and the angle of the tilt 
with which the \hii region is seen. The model is constrained by 
flux measurements of the source, temperature measurements of the 
gas, and the size and shape of the emission.

The contour maps of the model and the actual \hii region are shown 
in Figure~2. For presentation the contour map of W~3(OH) has been rotated 
by $30^\circ$ in position angle. The brightness along cuts made 
through the apparent major and minor axes are shown along the graph axes. 
In the model the electron temperature is assumed to be $10^4\,\rm K$ 
and the density $n_e=1.1\times 10^6\,\rm cm^{-3}$.
The shell minor axis radius is $0\secpoint 5$, the major axis radius, 
$0\secpoint 95$, and the thickness on this minor axis is $0\secpoint 13$. 
The angle between the plane of the sky and the major axis is $10^\circ$. 
The total mass in the ionized shell is $0.013\,M_\odot$.

The thickness of $0\secpoint 13$ along the minor axis means that on the major 
axis, the thickness is less than $0\secpoint 04$, or about 80~AU. 
This is a consequence of the assumption of constant density in the ionized
gas, and the ionization constraint could alternatively be satisfied
by a thicker shell of lower density gas.
However, a careful inspection of the brightness profile 
along the major axis in Figure~2 shows that there 
are small peaks at the edges of the emission region. This feature is 
reproduced in the model, and is a consequence of having a very thin shell. 
Another possible origin of the peaks is insufficient deconvolution 
of the maps. However, if this were the case, there would be negative
features of roughly the same magnitude around the source, which are
not present in the maps. It is likely that the actual shell
is incomplete with gaps in some areas, especially near the poles of the 
prolate shell. Towards the eastern limb there is evidence of an 
``champagne-type'' flow emerging, leading to a large area of low-emission 
level to the NE of the main source (Baudry et al. 1993; Keto et al. 1995), 
which is discussed in \S\ref{dynamics}.

The simple model described here cannot reproduce the bright notch on the 
northern limb, nor the tail of emission that extends towards the east. 
As Dreher \& Welch~(1981) noted, the \hii region is also clumpy, which 
becomes more evident in their 1.3~cm maps. The present model 
does not consider the lateral spreading of ionizing photons within the 
shell, but effects caused by this should be mostly washed out on the 
size scale of the beam. Finally, a careful inspection of W~3(OH) 
reveals many fine, wispy structures. No simple model can sufficiently
accommodate all such features. 

\subsection{Molecular environment surrounding the \hii region}

By combining the measurement of the angular expansion rate and the
density in the ionized shell determined from the model, we can estimate
the physical conditions in the region immediately bordering the
ionization front and in the ambient region ahead of the shock.
These estimates of the physical conditions may then be 
compared with observed values. For the purpose of simplicity, 
the ionized shell will be assumed to have spherical symmetry,
and we use parameters appropriate to the equatorial region of the
shell. This is adequate because the aspect ratio of the model is
$0.5:0.95$, which is less than the typical fractional uncertainty of 
observational measurements and is certainly less than the 
fractional uncertainties we expect from the estimates in
the following discussion. Furthermore, since the real conditions
are complicated by clumping and gradient effects, the 
estimates computed below should be taken to indicate average 
conditions near the \hii region.

We draw on an analysis which has been carried out by many workers 
(cf. Spitzer 1978). In this picture the \hii region is excited by
a central star with a constant output of ionizing photons. The \hii
region is a thin shell, which is radially expanding with a velocity $v_i$
with a uniform number density $n_i$. The \hii region is bordered by
a discontinuous ionization front, where the hot ionized gas meets
the shell of molecular material. This neutral shell is 
relatively thin, but since its bulk velocity is supersonic,
there is a shock where the shell impacts the ambient material. 

The total amount of material in the shell enclosed in a region less than a 
distance $r$ from the star is constant: $r^3n =\rm constant$. 
This implies that at the ionization front $3v_in=-r_i\dot n_e$, where 
the subscript $i$ indicates parameters for material that is immediately
next to the ionization front. In order to relate $v_i$ to the ionization 
front velocity, $V_i$, we recall that the total number of recombinations 
in the shell is also constant: $r_i^3n^2 = \rm constant$, 
where $r_i$ is the radius of the ionized region. This equation implies that 
$3V_in_e=-2r_i\dot n_e$ at the ionization front. Hence, the velocity of 
the ionized material can be related to the movement of the ionization front: 
$v_i={1\over 2}V_i$. The velocity of ionized gas with respect
the ionization front at rest is $u_i={1\over 2}V_i$.

It is then a simple matter to calculate the conditions across
the ionization front. First, we assume that the sound speed 
in the neutral region is negligible. The value of $u_i$ indicates 
that the ionization front is density-critical, and that the ram pressure
of the ionized material is insignificant compared to its
thermal pressure. Thus, the ratio of the density in the neutral shell 
to that in the ionized region may be expressed as $\rho_s/\rho_i=
c_s^2/c_i^2$, where $c_s$ and $c_i$ are the sound speed
in the neutral shell and ionized region, respectively. This expression
may be then expressed as $n_sT_s=n_iT_i$, where $n_s$ and $n_i$ are
the number density in the ionized and neutral regions. Here we assume
an isothermal gas, and ignore the pressure arising from the 
magnetic field and turbulence.

The temperature of the ionized gas is taken to be $T_i=T_e=10^4\,\rm K$,
which is typical of temperatures in \hii regions. The density of the
\hii region, taken from the model in \S\ref{modelsection}, is 
$n_i=1.1\times 10^6\,\rm cm^{-3}$, where we assumed that $n_i=n_e$. 
The physical conditions in the shocked shell may be inferred from 
observations of hydroxyl masers, which are believed to be harbored in the 
same dense shell of material adjacent to the  \hii region 
(Moran et al. 1968, Reid et al. 1980). The masers occur in warm, 
dense clumps of molecular material, with $T\sim 150\,\rm K$, 
$n\sim 10^{6-8}\,\rm cm^{-3}$ (Reid \& Moran 1988). From VLBI proper 
motion observations of these masers, Bloemhof et al. (1992) determined 
that they are also undergoing expansion at a speed of 
$\sim 3\,\rm km\,s^{-1}$, a value which is consistent with 
the observed expansion velocity of the ionization front, lending support to 
the claim that these masers occur in the expanding neutral shell. 
However, these masers occur in discrete clumps, with sizes
$d\sim 10^{14}\,\rm cm$ across, and hence the density range given
above are only characteristic of the conditions inside a masing
region, which is a very small fraction of the volume. Observations 
of highly excited transitions of OH, which probably only arise in the 
warm and dense environment that also fosters the OH masers, indicate 
that the average density is $n\sim 7\times 10^6\,\rm cm^{-3}$ 
and $T\sim 150\,\rm K$ (Cesaroni \& Walmsley 1991, Baudry et al. 1981). 
Taking $n_s=1\times 10^7\,\rm cm^{-3}$, we find that the pressures are
somewhat different, $n_sT_s=1\times 10^9\,\rm K\,cm^{-3}$ and 
$n_iT_i=1\times 10^{10}\,\rm K\,cm^{-3}$. The latter quantity 
suggests that the number density in the shocked shell should be
about $n_s=7\times 10^7\,\rm cm^{-3}$, which is the upper
limit to the density estimate given by Reid \& Moran (1988). 
Such a high density, however, is more favorable to the
formation of high-gain masers. We may be able to reconcile
this slight discrepancy, however, with the presence of a
magnetic field.

If the magnetic field is sufficiently strong, then it can play an 
important role in the dynamics of the \hii region and its immediate 
neighborhood. The effect of magnetic fields is particularly important
in strong shocks, because their presence limits the degree to which
shocked gas can be compressed. Also, if the Alfv\'en speed exceeds 
the sound speed, then the magnetic field can dominate the energy 
transfer through a medium. However, the measurements of magnetic
field strengths are difficult, and within the ionized region, such
measurements do not exist. Also, it is necessary to have some
information about the direction of the magnetic field with
respect to the ionization front, but this is not readily observed.
Therefore the conditions near the shock fronts cannot be modeled 
precisely. 

The measurements of the magnetic field near W~3(OH) have been
made by observing the Zeeman effect in hydroxyl masers.
At the positions of the OH masers the typical magnetic
field strength is $B_{tot}\sim 5\,\rm mG$ (Moran et al. 1968,
Reid et al. 1980). Recently, G\"usten, Fiebig \& Uchida~(1994) 
measured $3.1\pm 0.4\,\rm mG$ from observations of thermally excited 
OH. With this field strength, the magnetic field pressure in 
the gas is $B^2/8\pi\sim 6\times 10^{-7}\,\rm dyne\,cm^{-2}$, which is
higher than the thermal pressure in the neutral gas by an order 
of magnitude, $nkT\sim 2\times 10^{-8}\,\rm dyne\,cm^{-2}$. Furthermore, as
noted by Reid, Myers \& Bieging (1987), the thermal pressure
in the ionized material of the \hii region is nearly equal
to the magnetic pressure in the neutral region, suggesting
that the magnetic field must play an important role in the
dynamics of the \uchii region. However, they do not elaborate 
this assertion. Reid et al. (1987, Table 1) also tabulate
the thermal, magnetic, turbulent and ram pressure contributions
in the \hii region and molecular envelope.

To approximate the effect of the magnetic field at ionization front,
the ratio of densities can be approximated to first order by 
$n_s/n_i=v_A^2/c_i^2$, where we replace the sound speed
by $v_A$, the Alfv\'en speed. With this approximation, we see 
that $n_s/n_i\sim 4$, which is more consistent with the observed
data, but surely a very crude estimate. At the shock front
the magnetic field pressure dominates the thermal pressure
but is comparable to the ram pressure. Thus the effect of the
magnetic field is to push the ratio of the densities in the 
shell region and the ambient medium closer to unity.
Thus, although the observations do not permit us a detailed
analysis of the ionization front, it is clear that magnetic fields
play an important part.

Ahead of this shell of neutral material, there is a shock
as the shell impacts the ambient material. Making a similar analysis,
we find that at the shock front, $P_i=\rho_aV_s^2$, where $\rho_a$
is the density of the ambient gas. In this case most 
of the pressure is supplied by the ram pressure, because the
temperatures of the gases are not very different, and the sound
speed is much smaller than the expansion rate. This expression
predicts that the ambient number density to be $n_a/n_s=c_s^2/V_s^2=0.1$. 
The ambient material can be probed by observing molecular
lines in absorption against the \hii region, although with this
method some of the gas in the shocked region will be sampled as well. 
Absorption of $\rm H_2CO$ near the velocity of the OH masers indicates 
that the density of the foreground $\rm H_2$ number density ranges 
from $n<10^3\,\rm cm^{-3}$ towards the eastern limb to
$n>5\times 10^5\,\rm cm^{-3}$ in the west (Dickel \& Goss, 1987). 
This is consistent with the density observed in the regions 
harboring OH masers.

In our simple analysis, we have ignored the effects of turbulence, 
which has been proposed as a mechanism through which \uchii regions 
attain their longevity (Xie et al. 1996). The effects of 
turbulence is usually treated simply as an extra pressure term, 
$P_{turb}$. The pressure arising from turbulence is particularly important 
in the \hii region, where its value is comparable to the thermal pressure. 
This would mean that more pressure is needed from the shocked gas, but 
would change the value of the density by less than an order of magnitude. 

\subsection{Stellar wind and radiation pressure}

Soon after the central star in W~3(OH) reached the main-sequence, 
the stellar wind and radiation pressure drove away the circumstellar 
material to evacuate a cavity, and the \hii region was formed from
the inner edge of the shell of material (Davidson \& Harwit 1967).
Presumably, this is picture still applies to W~3(OH). 

In order to maintain the shell structure, and to keep it expanding, 
there has to be some pressure being exerted from inside the shell. 
The magnitude of the pressure necessary to support the shell
can be easily estimated by setting the thermal pressure in the
ionized region equal to the ram pressure in the rarefied
region, $P_{ram}=2n_ek_BT_e$. The two possible forces responsible 
for this pressure are the stellar wind composed of material 
from the star and radiation pressure acting upon dust in the shell.

It is well-known that early-type stars possess fast 
strong stellar winds, $v_w\sim 1000\,\rm km\,s^{-1}$ and associated 
mass loss, $\dot M\sim 10^{-6}\,M_\odot\,\rm yr^{-1}$ for an O7
star, although the nature of their mechanism is poorly understood 
(e.g. Chiosi \& Maeder 1986). If we assume that the mass
loss is isotropic, then the ram pressure of 
the stellar wind is $P_w=\dot M v_w/4\pi r^2$, where $r$ is the 
distance of the shell from the star. The wind pressure is
$P_w=2\times 10^{-6}\,\rm dyne\,cm^{-2}$, which is nearly
equal to the thermal pressure of the gas, $P_i=3\times 10^{-6}\,\rm
dyne\,cm^{-2}$. Thus we calculate there is sufficient ram pressure 
from the stellar wind to sustain the shell structure in W~3(OH). 

The radiation from the central star is first absorbed
by dust particles, which share their momentum with the
gas through collisions. Dreher \& Welch (1981) show
that about two-thirds of the ultra-violet radiation is
absorbed by dust in the ionized shell. 
The pressure from the dust can similarly be expressed as,
$P_{rad}=\beta L/4\pi r^2 c$, where $L$ is the luminosity of the
star, $c$ is the speed of light, and $\beta$ indicates the
fraction of the radiation that is absorbed by the dust,
which in this case is about unity. We calculate that 
$P_{rad}=3\times 10^{-6}\rm\,dyne\,cm^{-2}$, 
which is almost equal to the stellar wind pressure.

The outward pressures stated above represent merely lower
limits to the actual values. Nevertheless, it is quite clear
that there is sufficient pressure from either these
mechanism to evacuate W~3(OH) and to keep it expanding.
It is not possible to determine which process dominates
in evacuating the shell, although they probably work in
parallel.  

\subsection{A dynamical model for W~3(OH)}
\label{dynamics}

One of the most perplexing aspects of W~3(OH) is the systematic blue-shifting 
and broadening of hydrogen recombination lines with increasing quantum number.
Early observations of the region with centimeter-wave interferometers
revealed that there was a significant velocity difference between the 
foreground molecular material, probed by OH masers (Reid et al 1980) or
methanol absorption lines (Reid et al 1987), and that of the 
ionized region, probed by high-quantum number recombination
lines of hydrogen (Hughes \& Viner 1976). Reid et al (1987) 
measured a difference of about $5\,\rm km\,s^{-1}$, which they interpreted 
as infall of the molecular material towards the star. However, 
observations of recombination lines at higher frequencies (lower quantum
numbers) showed that the velocity difference becomes less. In a 
prescient paper, Berulis \& Ershov (1983) attributed this trend
to non-LTE effects in a rapidly expanding ionized shell.
The shell was thought to be expanding at $18\,\rm km\,s^{-1}$.
Further study was permitted by the advent of millimeter-wave 
interferometers, and Welch \& Marr (1987) showed that the velocities 
of the recombination lines at lower quantum numbers closely approach 
those of molecular lines. They concluded that the infall model was wrong, 
and also the line width of the recombination line was too narrow for the
source to be rapidly expanding. 

The present observations unambiguously preclude the rapidly expanding 
shell model of Berulis \& Ershov~(1983), unless the expansion only
occurs along the viewing axis, which seems highly unlikely.
More recent attempts to describe W~3(OH) are the cometary
stellar-wind bow shock (Van Buren et al. 1990), the line-broadened
fast flow (Keto et al. 1995), and the two layer cloud model 
(Wilson et al. 1991).

Our expansion measurements do not support the suggestion by Bloemhof
et al. (1992) that W~3(OH) is a cometary stellar-wind bow shock 
(Van Buren et al. 1990). The cometary bow shock 
model was specifically proposed to explain the longevity of \uchii regions. 
In the bow shock model the central star has a large velocity relative to 
the surrounding molecular cloud, and the stellar wind pushes aside
the oncoming molecular material. The ram pressure of the swept-up molecular 
material effectively confines the \hii region, and its age is thus 
considerably longer 
than indicated by the dynamical age. Furthermore, the cometary phase 
may last as long as there is sufficient material near the star: as it
traverses different molecular cloud clumps, the cometary bow shock may
reform a number of times. Most relevant to our observations, 
the bow shock structure can only appear to expand if the star is moving
from a more dense to a less dense region of the molecular cloud. We make
a simple estimate of the required density gradient as follows.

In the simple stellar-wind bow shock model an early type star is assumed
to plow through a molecular cloud at a supersonic speed of about 
$6\,\rm km\,s^{-1}$. Since Welch \& Marr (1987) determined that the relative
velocity along the line of sight between the W~3(OH) \hii region and
the molecular cloud is less than $1.3\,\rm km\,s^{-1}$, the hypothetical
velocity of $6\,\rm km\,s^{-1}$ must occur in the plane of the sky,
possibly to the SW (Bloemhof et al 1992). The radius, $l$, of the bow
shock is determined by pressure balance between the stellar wind
and the ram pressure, and is given by $l\sim n^{-1/2}$ (van Buren et al 1990).
The apparent expansion of the \hii region owing to density variations 
in the molecular cloud may be written as $\epsilon = -d(\ln n_a)/dr(v_s/2)$, 
where $1+\epsilon$ is the expansion factor, $n_a$ is the number density 
of the ambient medium. By simply solving for the gradient, we compute 
the lower limit to the density gradient: $d(\ln n_a)/dr>140\,\rm pc^{-1}$. 
Given the size of W~3(OH), we estimate that density has to be falling off 
by a factor of at least 4 from NE to SW across the \hii region. This is 
in direct contradiction to what is generally observed using various 
molecular tracers, where the gradient goes in the opposite 
sense (e.g., Dickel \& Goss 1987). Furthermore, if there 
were such a gradient to begin with it is very difficult to believe 
that W~3(OH) would be presently so symmetrical, since the radius
should change by a factor of 2 across the region.

Keto et al. (1995) modeled the W~3(OH) as a fast flow of ionized
gas moving towards the observer that included the effects of
pressure broadening in the gas. The flow is densest at its origin,
nearest the star, and becomes less dense further along the flow.
The flow fills a paraboloid, and
the velocity of the material is fixed to conserve mass.
For a viewer looking down the flow, the recombination line
from the gas to the foreground is optically thin, blue-shifted, and
has a thermal line width. However, at the densest part of the 
flow, the same recombination line is pressure-broadened and 
is effectively made optically thin. (The opacity integrated
over a large velocity range, however, is still proportional 
to $n_e^2$.) The effects of pressure
broadening, however, decreases at lower quantum numbers. Therefore,
the millimeter-wavelength recombination lines will appear
thermally broadened and centered at about the same frequency
as the molecular gas. At lower frequencies the line
becomes broader and more blue-shifted. Keto et al.~(1995) also try 
to account for the observed shell 
structure using the flow. Specifically, they tilt the parabolic flow so
that the flow resembles the shape of W~3(OH), with the opening
pointing towards the northeast. The tilted parabolic flow qualitatively
describes both the observed structure and the velocity gradient in the
recombination line across the major axis of the \hii region.
Finally, to account for the limb brightening, the inner one-sixth of 
the parabola is left without any material flowing through it. 

Although the model gives good agreement to the line width and 
line velocity data, the only physical constraint in
the model is mass conservation. Furthermore, Keto et al. (1995) 
do not propose any physical mechanism responsible 
for the flow other than to invoke the well-known champagne flow model of
Tenorio-Tagle (1979). There are several salient problems even in the 
simple model they present. First, when the emission measure 
is calculated from the parameters they give for the flow, it falls short 
of the observed value. An upper limit to the emission measure may
be calculated as 
$EM=n_0^2z_0/(2\alpha_{ne}-1)\approx 2\times 10^7\,\rm cm^{-6}\,pc$,
where we use the numbers provided by Keto et al. (1995) and use their
notation. Here, $n_0$ is the maximum density, $z_0$ describes the 
size scale of the \hii  region, and $\alpha_{ne}$ is the exponent by 
which the density in the region falls off in units of $z_0$. 
We assume that $z_0$, which is not mentioned in their paper,
is the size of the \hii region itself. The actual EM is about 50 
times higher. Thus the flow they describe is optically thin.

Since the flow is optically thin, we might consider a model in which 
this parabolic flow is actually emerging from the optically thick shell 
shell. This description is somewhat similar to that given
by Wilson et al. (1991) who modeled the \hii region as being composed 
of two discrete layers of ionized gas moving at different
velocities. In this configuration the observed recombination lines 
would mainly trace the rarefied fast flow because those that 
originate in the shell are severely pressure-broadened.
As noted in \S\ref{modelsection}, there is extended emission 
towards the NE edge of W~3(OH), and we can picture that the fast 
flow is emerging from the optically thick shell to the foreground. 

In this configuration the apparent expansion velocity of shell would
be slightly higher than that of the actual bulk motion of the ionized 
gas, since the radiation can reach further into the neutral medium 
as the region gets evacuated. We establish that in order to maintain 
such a flow for an extended period of time, the ionized material must 
continually be replenished. The time scale for the flow to 
evacuate the ionized region without replenishment is 
given by $t_{evac}=M/\dot M\sim 1000\,\rm yr$, where $M$ 
is the mass of the ionized region, and $\dot M=vnAm_H$ is
the mass flux. In the expression for the mass flux $v$ is the
velocity of the flow, $n$ is the number density of the flow, $A$ is the
cross-sectional area, and $m_H$ is the proton mass.
Here, we assumed that the scaling length $z_0$ in Keto et al. (1995)
is the about the same size as the \hii region,
$\sim 2\times 10^{16}\,\rm cm$. If the material is replenished from
ionization front, then there should be an apparent movement
of the ionization front. We can estimate this movement as
$v\sim\dot M/ 4\pi r^2n$, where $r$ is the radius of the
shell. We estimate that $v\sim 2\,\rm km\,s^{-1}$. 
If the depletion of gas is contributing to the movement of
the ionization front, then the masers would appear to expand at a
slightly slower speed than the ionization because the bulk movement of
the gas would still be determined by radiation or stellar wind
pressure. The ionized region does seem to be expanding slightly
faster than the masers, although the uncertainties are 
rather too high to make a decisive conclusion.

We should note that the fast flow model does not adequately agree
with several important observational facts. Keto et al. (1995) 
observed that the $\rm H92\alpha$ line velocity has a gradient 
of $18\,\rm km\,s^{-1}$ from E to W. When the velocity gradient 
across the source is calculated from the model, it cannot easily reproduce 
the observed gradual shift in the recombination line velocity across the 
source. Furthermore, it may be merely a coincidence that the major
axis of W~3(OH) is in the same direction as the velocity gradient.
For example, in the case of G34.3, another bright, well-studied
\uchii  region, the gradient in the velocity of the recombination line
is transverse to the long axis of the \uchii region (Gaume et al. 1994).
The model also makes definite predictions about the lineshapes. 
The line profiles of millimeter-wave recombination lines
have a blue-shifted shoulder because while the bulk of the emission
originates at the densest area, the higher velocity material 
still makes a contribution. Unfortunately, the line profile of 
$\rm H35\alpha$ does not show even a hint of asymmetry 
(Wilson et al. 1987). The centimeter-wave line profiles should also 
be slightly asymmetric, with the shoulder on the low velocity side. 

These discrepancies might lead us to conclude that rather than
there being a fast flow, there might be layers of ionized
gas moving more or less at discrete but uniform speed in the 
foreground of the main \hii region (Wilson et al. 1991). 
Wilson et al. (1991) modeled the \hii region as comprised of two 
components along the line of sight. The first clump has a higher 
density but smaller mass than the second clump, and the observer 
sees a blend of the two regions. With this configuration, 
Wilson et al. (1991) were able to reproduce the line width and 
line velocity data, and it seems that this model would readily 
reproduce the observed profiles of the recombination lines.

Thus, we come to the general conclusion that that the most 
satisfactory model for W~3(OH) is that it is comprised of two 
components, an optically thick shell and a related or separate 
regions of ionized gas in the foreground that are moving at 
different velocities. Ultimately, further constraints to models 
will have to come from better measurements of the recombination
line profiles. The measurement of Wilson et al. (1987) favors 
the discrete cloud model. 

\subsection{Lifetime of the \uchii phase}

The direct measurement of the expansion rate in W~3(OH) gives us
a good estimate of the age of the \hii region that is independent of
any models: The age of the shell is only $2.3\times 10^3\,\rm yr$.
This assumes that the expansion velocity has been constant
in time, which, of course, is a simplified picture. If we linearly 
extrapolate to the future, then W~3(OH) would cease to become an 
ultra-compact \hii region in about $2\times 10^4\,\rm yr$. This 
is within an order of magnitude of the expected duration of the 
ultra-compact phase.

In another simple analysis we can apply the classical equation 
that governs the expansion rate of an \hii region once it has
reached the Str\"omgren radius. This was first applied to the
study of \uchii regions by De Pree, Rodr\'\i guez \& Goss (1995),
who proposed that if the molecular gas in which the \hii region 
initially formed is sufficiently dense and warm, the ambient pressure 
may be sufficient to confine the \hii region for at least $10^5\,\rm yr$. 
We take their example of an O6 star which formed in an ambient
density of $n=2\times 10^7\,\rm cm^{-3}$. These conditions are
reasonable since they are close to the values used for the static
model for the shell. The initial Str\"ogrem sphere is only 
$10^{-3}\,\rm pc$ in radius, and the initial rate of expansion is
at the plasma sound speed. To expand to a diameter of 0.01~pc, which
is the present size of W~3(OH), it takes only 900~yr, but by this time, 
the rate of expansion has slowed to $4\,\rm km\,s^{-1}$. This value 
for the expansion rate is in remarkable agreement with our measured 
value. In this analysis the expansion rate continues to decrease with 
time, and at the end of $10^5\,\rm yr$ the \hii region reaches a diameter
of only $0.14\,\rm pc$. Therefore, in the context of this picture
although the \hii region is relatively young, W~3(OH) can apparently 
remain in the \uchii phase for almost $10^5\,\rm yr$.

In summary it appears that the major properties of the W~3(OH) \hii
region can be satisfactorily explained by invoking simple, 
classical ideas regarding the Str\"ogrem sphere. Though the \hii region
is probably be very young, the predicted time it will remain in the 
\uchii phase is comparably long to what is expected. The complete 
picture, is obviously more comlicated than this. For example, since 
the density gradient would favor faster expansion rates as the 
region gets larger, the extrapolated lifetimes 
from the simple analyses might be considered an upper limit. The 
uncertainties regarding the molecular environment and stellar wind, 
radiation, and magnetic pressure make it difficult to predict very 
accurately the expected life time of the ultra-compact phase. 

Finally, it is important to distinguish the age of the structure
of an \hii region from the actual duration of the ultra-compact phase: 
the two are not necessarily related. The appearance of the ionized
gas may change quite rapidly. Thus, the dynamical age may not indicate 
the actual age of the \hii region at all. Indeed, in W~3(OH), the 
shell structure may well disappear in a rather short period of time, 
if the shell breaks out of the core of the molecular cloud;
it would then take on a different appearance, perhaps a core-halo 
structure. Hence, it will still be an \uchii region. Since there 
is evidence of activity near the central star (\S\ref{central_star}) 
this is a tantalizing possibility, although it is merely a speculation. 

\subsection{Emission near the central star}
\label{central_star}

As we noted earlier, there is a feature very close to the
projected center of the \hii region that exhibits rapid
variability, and owing to its location we tentatively link
to the central star. If the change in flux at this source
can be entirely attributed to mass loss from the region,
we infer that the density in the region fell by 
$\sim 2\times 10^5\,\rm cm^{-3}$. If this mass left the
region enclosed by $\sim 300\,\rm AU$ around the star within 
a period of $<4\,\rm yr$, this represents a mass loss of 
$\sim 10^{-6}M_{\odot}\,\rm yr^{-1}$, which is a value remarkably
close to the wind mass loss rate of the star. 

Curiously, this value for the mass loss is of the same 
order of the mass flux expected from the photoevaporation disk 
model (Hollenbach et al. 1994), and the size scale is also roughly 
the size of most disk models as well. Furthermore, another 
approximation can be made: because there are positive features beside 
the negative feature, a crude velocity may be estimated if we interpret
these features to be movement of gas material. The positive features are
roughly $0\secpoint 1$ from the negative peak. 
If this represents a movement of gas, it would correspond to 
a velocity of $\sim 300\,\rm km\,s^{-1}$. This 
is the characteristic speed of a stellar jet. 

\section{Conclusion}

We have made direct measurements showing that the \uchii region W~3(OH)
is expanding at $\rm 3\,\,to\,\,5\,km\,s^{-1}$. The expansion rate also 
implies directly that the age is $\sim 2300\,\rm yr$, much lower than the
$10^5\,\rm yr$ believed to be typical for \uchii regions. The new data 
are consistent with a simple physical model in which the OH masers are 
in a dense, shocked neutral shell swept up around the \uchii region,
which is expanding anisotropically into the surrounding molecular
cloud. Within this simple model there is apparently no difficulty
with the age issue regarding \uchii regions. In addition, from the 
difference maps we have detected a number of interesting sources, most 
notably the variability linked to the central star. There is a clear need 
to study this particular source and other flickering objects in greater 
detail.

The technique of generating difference maps could be applied to a few 
of the brightest shell or cometary \uchii regions. It was recently 
applied to observe an apparent expansion in G5.89$-$0.4, a 
well-studied bright shell-shaped \uchii region (Acord, Churchwell 
\& Wood 1998). Acord et al (1998) concluded that the \hii region 
is expanding at $35\,\rm km\,s^{-1}$. It should be stressed that 
the expansion signature can be best translated into a velocity 
when there are sharp features in the maps, like those found in an
ionization-bounded nebula. G5.89 has an extended region of emission around
the limb-brightened shell, and this makes the proper motion measurements 
somewhat ambiguous. However, an important result from Acord et al (1998) 
is that the time-scale for gross change in the emission structure of 
the \hii region is of order less than 100~yr. We conclude that W~3(OH)
is quite distinct from G5.89, despite their similar morphology. 
Regardless, this short time-scale clearly has important implications 
for the study of \uchii in general. We have already tried the 
same experiment on a core-halo region, NGC~7538 (Kawamura 1997). 
Although large differences were observed between maps at different 
epochs, there was no clear pattern and it was hard to make a simple 
physical model.

We thank the anonymous referee for suggesting that we expand our
discussions regarding the cometary bow-shock model and the lifetime
issue.

\end{document}